# Femtosecond Photonic Viral Inactivation Probed Using Solid-State Nanopores

Mina Nazari, [1,4] Xiaoqing Li, [9] Mohammad Amin Alibakhshi, [7] Haojie Yang, [10]
Kathleen Souza, [8] Christopher Gillespie,[⊥ 8] Suryaram Gummuluru,[6] Björn M. Reinhard, [3,4] Kirill S. Korolev, [2,5]
Lawrence D. Ziegler,[3,4] Qing Zhao, [9] Meni Wanunu,[7*] Shyamsunder Erramilli, [2,4*]

*Departments of [1]Electrical and Computer Engineering, [2]Physics, [3]Chemistry and [4]The Photonics Center, [5]Bioinformatics Program, Boston University, Boston, MA 02115, United States*
*[6]Department of Microbiology, Boston University School of Medicine, Boston, MA 02118, United States*
*[7]Department of Physics, Northeastern University, Boston, MA 02115, United States*
*[8]Next Generation Bioprocessing, MilliporeSigma, Bedford, MA 01730, United States*
*[9]School of Physics, Peking University, Beijing, P. R. China*
*[10] Department of Mechanics, Southeast University, Nanjing, China.*

We report on the detection of inactivation of virus particles using femtosecond laser radiation by measuring the conductance of a solid state nanopore designed for detecting single virus particles. Conventional methods of assaying for viral inactivation based on plaque forming assays require 24-48 hours for bacterial growth. Nanopore conductance measurements provide information on morphological changes at a single virion level. We show that analysis of a time series of nanopore conductance can quantify the detection of inactivation, requiring only a few minutes from collection to analysis. Morphological changes were verified by Dynamic Light Scattering (DLS). Statistical analysis maximizing the information entropy provides a measure of the Log-reduction value. Taken together, our work provides a rapid method for assaying viral inactivation with femtosecond lasers using solid-state nanopores.

## I. INTRODUCTION

Existing and emerging viruses are a major threat to human and veterinary public health. The need for safe and reliable inactivation or removal of viruses is universal in antiviral therapies, pharmaceuticals, and viral vaccine development. Conventional pharmaceutical pathogen inactivation methods are quite effective, but they involve substantial collateral damage and have undesirable side effects. [1-3] Chemical-free viral inactivation methods such as ultraviolet (UV) and gamma-irradiation have been used to minimize some of the side-effects. Unfortunately, these methods still adversely affect thermolabile compounds and denature biomolecules of interest in the medium containing the virus. Ultrashort Pulsed lasers (USP) provide new opportunities for chemical-free pathogen disinfection in solution. Photonic methods have the potential to provide an attractive alternative to existing biocides and ionizing radiation techniques. [4-8] Photonic inactivation has been successfully achieved with the focused femtosecond (fs) laser pulses for exposure times of ≥ 1h on sample volumes typically of ≤2 ml. [4-7,9,10] Although the ultrafast laser inactivation method for viral inactivation is fairly well established, a systematic understanding of the inactivation mechanism, which can contribute to the design and optimization of protocols, is currently lacking.

Viral inactivation is a complicated process and its outcome highly depends on the specific treatment method. A wide range of biological assays could detect and quantify intact viruses in an ensemble manner which is extremely laborious, time-consuming, and low-sensitive. [11] A different approach is to explore viruses at the single virion level. Different optical methods have been developed to characterize single virus; [12-14] but still there is a need for a technique that is fast, sensitive and uses low sample volumes. Although imaging techniques, such as AFM and TEM, are capable of characterizing viruses with high sensitivity, results will be inevitably affected by the tedious and costly sample preparation steps. Nanometer-sized pores in a membrane offer the capability of electrically detecting molecules in a label-free manner at single-molecule level in a volume as small as a few microliters and detection times as short as a few seconds. Passage of molecules and particles through a nanopore causes transient disruption in the ion current flux through it, from which the size, concentration, and distribution of analytes can be deduced. [15,16] The electrical signal characteristics in a given analyte sample

---

[⊥] *Current address of C. Gillespie:* Immunogen, 830 Winter St Waltham MA

[*] wanunu@neu.edu and shyam@bu.edu

strongly depend on the analyte passing through the pore, the pore geometry, and the experimental conditions such as pH, ionic strength, applied voltage, temperature, etc. This single-particle electrical sensor has been used for quantifying the conformational properties of proteins, [17-23] understanding DNA transport [24-26] and detecting small molecules, [27,28] among many other applications. Previous studies have reported the ability of nanopore sensors to detect spherical and icosahedral viruses,[29] virus capsids, [30] the masses and zeta potentials of viruses, [31,32] and to explore the translocation of a stiff, rod-shaped virus. [33]

This ability of nanopores to detect the translocation of nanometer size particles motivated us to study the effect of an optical viral therapy on a single virus level, crucial for the preparation of very safe biotherapeutics. In this paper, high incident fs laser intensities of >100 GW/cm$^2$, which are more than $10^5$ times greater than the previous studies, have been used to inactivate viruses, leading to 4-log reduction in viral activity reduction in 1 min irradiation of ~ 2 ml sample volume. This result shows nearly more than four orders of magnitude improvement in treatment time compared to conventional pulsed laser viral inactivation methods.[34] Furthermore, we demonstrate the capability of the nanopore technique to precisely characterize individual viruses, explore how vital viral function is affected by treatment, and quantify how effective this label-free viral inactivation technique is. In light of these points, we investigate the effects of fs laser on inactivated ΦX174 bacteriophage, which has the first sequenced DNA-based phage genome widely used standard for viral clearance, as well as a surrogate for enteric human viruses. [35] By electrically counting virus particles in a small volume sample, we monitor here changes in the physical properties of treated virus samples. Further, we develop a statistics-based method, to monitor the reduction value of viruses using sequential measurements, and compare it with a plaque-forming assay. Moreover, we compare the effect of inactivation on viruses in an ensemble manner and at the single-virus level using Dynamic Light Scattering (DLS) measurement and nanopores, respectively.

## II. METHODS

**Femtosecond laser irradiation.** A femtosecond laser based upon a Legend Elite Duo (Coherent Inc.) Ti-sapphire regenerative amplifier has been used as the excitation source in this study. The laser produces a continuous train of 35 fs pulses at a repetition rate of 1 kHz. The output of the second harmonic generation system of the laser centered at ~400 nm with energies up to 2.5 mJ was used to irradiate the virus samples. Figure 1a depicts the experimental setup. The laser beam with spot size ~1 cm$^2$ was incident upon a typically 1 cm quartz cuvette containing 2 ml of virus sample while a stirrer was used to homogenize the virus's interaction with the laser beam. For ΦX174 sample with $10^{12}$ pfu/ml concentration, the laser treatment is made by exposing 250 μl of viruses in 2 mm cuvette. Typical sample exposure times to the laser beam were 15 min. All experiments were carried out at 22 °C, and all samples were immediately stored at 4°C after irradiation. All experiments have been done in triplicate.

**Virus sample preparation.** ΦX174 samples with $2 \times 10^{12}$ plaque forming unit (pfu) per ml concentration (Promega TiterMax ΦX174 Bacteriophage) in 0.05 M Sodium Tetraborate were stored at -80°C. Before the experiment, samples were thawed to room temperature, aliquoted, and kept at 4°C. For diluted samples, ΦX174 spiked feed solutions were prepared by serial dilution in Sorenson's buffer to the final concentration of approximately $10^6$ pfu/ml.

**Infectious plaque assay.** To count the ΦX174 in the solution, samples were diluted with Sorensen's buffer dilution blanks, to bring plaque to within a statistically valid range of 30-300 plaques per plate. Samples were assayed in triplicate by adding 0.1 mL of diluted sample and 0.1 mL of host cell suspension to a test tube containing 3 mL of molten (46-48 °C) ΦX174 overlay agar consist of 10 g of tryptone peptone (Difco), 8.5 g of agar (Difco) and 5 g of NaCl per liter of reagent water. Then the solution containing host cell and bacteriophage was vortexed and transferred to the ΦX174 bottom plate agar (2.5 g of NaCl, 2.5 g of KCl, 10 g of tryptone peptone from Difco, 10 g of agar from Difco, and 1 mL of 1 M $CaCl_2$ per liter of reagent grade water) and incubated overnight at 37 °C. Then plaques were counted and the corresponding bacteriophage concentrations were reported as pfu/mL. Sorensen's phosphate buffer (pH 7.3), ΦX174 bottom plate agar, and ΦX174 overlay agar were purchased from Northeast Laboratory Services (Winslow, ME).

**Nanopore device fabrication and measurement.** Our electrical detection system is composed of a nanopore formed in a 50-nm-thick insulating silicon nitride (SiN) membrane. The silicon nitride is deposited on a silicon substrate with 2-micron silicon dioxide previously grown on, which is chemically etched by potassium hydroxide (KOH) to obtain a freestanding membrane. The electron beam of a JEOL 2010F transmission electron microscope (TEM) was finely focused on the membrane in order to make a pore with controlled size. [26] The nanopore chip is then mounted in a fluoropolymer cell that allows electrical measurement of ionic current through the nanopore. The cell is filled with 0.1 M KCl solution (16.1 mS/cm conductivity), buffered to pH 7 using 10 mM tris. The silver-silver chloride (Ag/AgCl) electrodes are inserted in both cis- and trans- chambers, and a DC voltage is applied to flow current and drive charged molecules across the pore.

**Data acquisition and analysis.** A Chimera VC100 (Chimera Instruments LLC) was used for recording the ion current through the nanopore. Data was digitized at 4.17 MS/s, and streamed/saved to the computer at a 1 MHz bandwidth. Prior to analysis, recordings were further filtered using a 100 kHz digital low-pass filter. To verify the pore's stability, before the introduction of a virus sample to the nanopore, several seconds of current were collected to ensure that no event is detected and the baseline current is stable. Three key independent parameters are extracted from the nanopore data: the dwell time of viruses at the pore, $t_d$, the fractional current blockade, $F_I$, and the inter-event waiting time, $\delta t$, from which virus capture rates can be extracted.

**Dynamic Light Scattering.** In order to obtain information on the size distribution of viruses in an ensemble manner, we performed DLS measurement using Zetasizer Nano S90 from Malvern Corp. This DLS measurement is based on the Brownian motion of spherical particles; using the Stokes-Einstein relation the size is calculated based on measured diffusion constant of particles. [36] For the size measurement, ΦX174 virus with $10^{12}$ pfu/ml concentration and ΦX174 Virion DNA (New England Biolabs) with 1,000 μg/ml concentration, are diluted by 8-fold and 30-fold respectively in 10mM Tris. An aliquot of about 80 μl of the diluted samples is transferred to the cuvette for DLS analysis with the measurements done at 23 °C.

### III. RESULTS AND DISCUSSION

Obtaining extremely high levels of viral clearance is a substantial step in the purification of protein-based therapeutics. The presence of even a single virus in the final drug product could be harmful to the consumer's health. To prevent this, implementation of an effective viral inactivation strategy is crucial. USP viral inactivation with greater than 4-log reduction in viral infectivity would enable new chemical free pathogen clearance technology. [34,37,38] The first objective of this study is to extend the work done on USP inactivation of viruses, using a regeneratively amplified laser system. We expedite the reported USP photonic inactivation (>1 hr) [34] to 1 min. In this study, 35 femtosecond pulsed laser irradiation working at ~ 400 nm is used to irradiate 2 ml of ΦX174 bacteriophage for different irradiation times. Inactivation of viruses is measured by virus infectivity assay and the concentrations of the original samples were calculated by multiplying the plate count by the dilution factor, reported as pfu/mL. [39]

The final results for the viral inactivation experiments is reported as the Log Reduction Value (LRV) which provides a direct measure of viral inactivation. The LRV was calculated according to:

$$LRV = \log_{10}\left(\frac{C_U}{C_T}\right) \quad (1)$$

Here $C_U$ is the concentration of the untreated sample and $C_T$ the concentration of the treated samples exposed to the laser irradiation as described. Control samples consist of a sample with no laser exposure which was held under refrigeration during the experiment and another sample which experienced the same pipetting and stirring condition as treated sample but without the laser exposure. These control samples never differed significantly and were taken to check for loss of titer in the treated suspension. As represented in Fig. 1b more than $10^4$ reduction values of ΦX174 with $10^6$ pfu/ml concentration is achieved by irradiating 2 ml of virus suspension with 2.5 W femtosecond laser for different treatment times ranging from 1 min to 15 min. The combination of the large laser beam diameter with intensified beam results in fast viral treatment which can overcome the need for long irradiation times, and remove constraints on the corresponding practical implementation.

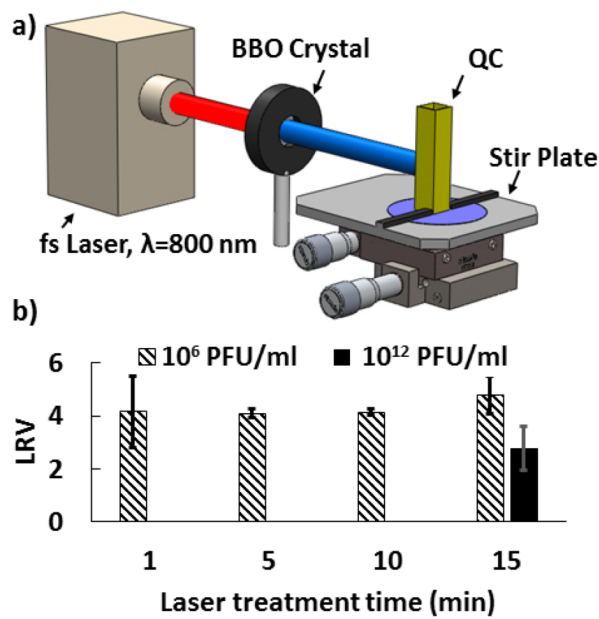

FIG. 1. a) Schematic of experimental setup for photonic viral inactivation. The following abbreviation are used: fs: femtosecond, QC: quartz cuvette, BBO: Beta barium borate. b) Log Reduction Value (LRV) measured for ΦX174 virus with $10^6$ pfu/ml (patterned) and six-orders of magnitude increased concentration (solid) after USP laser irradiation for different exposure time.

The next goal is to precisely monitor changes implied on the treated virus on the single virus level. A 20 μl aliquot of ΦX174 with ~$10^{12}$ pfu/ml is treated for 15 min exposure with the same laser setup. For this low volume of sample, we used a micro quartz cuvette with no stirring. Again, the strong reduction in viral infectivity (LRV>3) was achieved for six-orders of magnitude increased virus concentration (Fig. 1b).

As shown in Fig. 2, the ΦX174 viruses are being voltage-driven through a ~38 nm pores made of silicon nitride (SiN). A TEM image of the pore is shown as an inset. The pore size is intentionally chosen close to the virus size to slow down the translocations and allow for accurate measurement of the events. Viruses are electroosmotically driven through the nanopores upon application of a negative bias to the trans chamber. The application of voltage results in a steady-state countercurrent of $K^+$ and $Cl^-$ ions across the pore, which produces a stable baseline open pore current, $I_0$. When 0.5 μL of virus sample is added to 50 μL of buffer in the cis chamber, passage of individual viruses through the pore reduces the ionic current, which results in a spike in the current. This volume was sufficient to generate $> 2\times10^3$ events in 10 s for good statistics in untreated sample. The spike contains information about single virus, which can be extracted using statistical analysis methods. The current is measured by a Chimera VC100 amplifier which streams 1 MHz bandwidth data to a computer at a sampling rate of 4.17 MHz, followed by application of a 100 kHz low-pass filter in software to reduce the high frequency noise dominated by the chip capacitance.

Figure 3a shows continuous current traces obtained when untreated ΦX174 viruses were added to the cis chamber at 40 mV applied voltage. Each spike corresponds to transport of a single virus through the ~38 nm pore. The electric signal represents two predominant current levels, the higher one corresponds to open pore current when viruses do not translocate through the pore, $I_0$, and the lower corresponds to virus-occupied level. Based on the measured open pore current which is 1.13 nA, the pore conductance was calculated as $G=28.2$ nS. The inset shows a magnified view of a randomly selected event corresponding to individual intact ΦX174. The translocation current, $\Delta I$, which is the difference between the baseline current and the pulse minimum, depends on pore geometry and virus size. Knowing the nanopore diameter as 38 nm, the nanopore length was estimated from the open pore conductance to be $h_{eff} = 35\ nm$. This value, thinner than the overall initial 50 nm membrane thickness, $h$, was still thicker than the expected thickness of $h_{eff} = h/3$ previously found for the very narrow pores. [40] As shown in Fig. 3d, the mean fractional current blockade for untreated ΦX174 virus is measured to be 27%. This number is consistent with the 29% blockade predicted by the following theoretical equation, derived in Supplemental material, for a 30 nm diameter spherical virus:

$$F_I = \frac{\langle \Delta I \rangle}{I_o} = \frac{R_b - R_o}{R_b} \quad (2a)$$

In which $R_o = \dfrac{1}{\sigma d_p} + 4\dfrac{h_{eff}}{\sigma \pi d_p^2}$ is the open pore resistance and $R_b$ is the nanopore resistance in the blocked state:

$$R_b = \frac{1}{\sigma d_p} + 4\frac{h_{eff}-d}{\sigma \pi d_p^2} \quad (2b)$$
$$+ \frac{4}{\sigma \pi \sqrt{d_p^2-d^2}} \tan^{-1}\left(\frac{d}{\sqrt{d_p^2-d^2}}\right) \quad (h_{eff} \geq d)$$

Here $d$ is the virus diameter, $d_p$ is the pore diameter, and $\sigma$ is the salt conductivity which for the 100 mM KCl buffer was measured as 16.7 mS/cm using the conductivity meter.

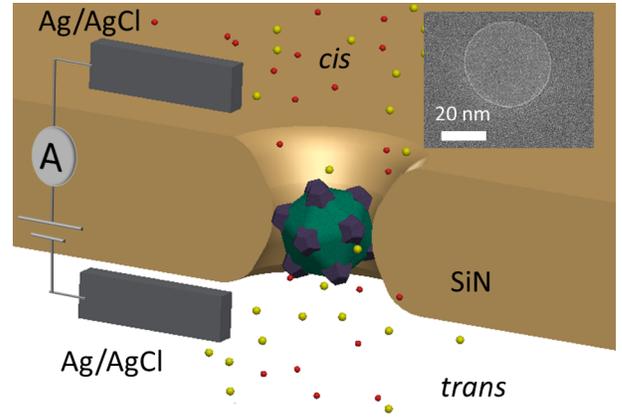

FIG. 2. Schematic of the nanopore setup for virus detection. Pores are fabricated in freestanding SiN membranes and an external bias is applied across the membrane to drive ΦX174 viruses, potassium ions (yellow dots) and chloride (red dots) ions through the pore. Inset: TEM image of a fabricated nanopore. (scale bar: 20 nm)

The agreement between theoretical and experimental translocation ratio verifies that the untreated viruses maintain their shape integrity during transport through the pore. Additionally, a 1-D drift-diffusion model can be used to describe the dwell time distribution of virus translocation through the nanopores. Fittings the probability density function (PDF) of dwell time distribution ($t_d$) with the following equation yields two important parameters: diffusion constant of viruses inside the nanopore, $D_{pore}$, and their drift velocity, $v_d$.

$$P(t) = \frac{h_{eff}}{\sqrt{4\pi D_{pore} t_d^3}} \exp\left(\frac{-(h_{eff} - v_d t_d)^2}{4 t_d D_{pore}}\right) \quad (3)$$

Application of this equation, which is based on the assumption of barrier-free transport, [41] for the untreated viruses (Fig. 3e) yields $D_{pore} = 0.5\,\mu m^2/s$ and $v_d < 10^{-4}\,m/s$.

To gain insight into the virus transport kinetics through this nanopore, we compare the in-pore diffusion coefficient of viruses with a bulk diffusion coefficient, using Stokes−Einstein equation $D = \dfrac{k_B T}{3\pi\eta d}$ where $k_B$ is the Boltzmann constant, $T$ the absolute temperature, $\eta$ the viscosity of the solution, and $d$ the hydrodynamic diameter of the virion. Using DLS we measured average diameters for ΦX174 of ~30 nm at 23°C, and accordingly, $D_{bulk}$ was calculated as $14\,\mu m^2/s$. The result shows small reduction in $D_{pore}$ which can be the consequence of virus-pore hydrodynamic interaction, as previously observed with protein transport through smaller pores. [42]

Upon successful detection of untreated viruses using nanopores, we probed the effect of laser treatment on the virus sample. A time trace for translocation of laser treated ΦX174 with LRV>3 (Fig. 1b) is shown in Fig. 3b. There is a drastic change in the time trace of treated viruses compared to untreated one which will be explored in more detail.

To explore the effect of laser treatment, the scatter plot of fractional current blockade versus dwell-time of treated and untreated viruses at 40 mv voltage is shown in Fig. 3c. Two clear groupings of events are visible noted, corresponding to the treated and untreated viruses which can be visually distinguished by drawing a line as shown in Fig. 3c. A histogram of fractional current blockades and dwell-time distributions for both treated and untreated samples at 40 mv along with generalized extreme value distribution fits to the distributions are shown in Fig. 3 d-e respectively. The untreated sample is centered at $F_I$ = 25.04+/-3.1 % with $\log_{10}(t_d)$ = 2.05+/-0.36 μs and the treated sample is centered at $F_I$ = 10.4+/-2.7 % with $\log_{10}(t_d)$ =1.3+/-0.12 μs. These two sets of independent parameters ($F_I$ and $t_d$) clearly show the effect implied on viruses by laser treatment. The $F_I$ is related to the size of the particle and hence a strong decrease in the fractional blockade yields the first important information on the effect of fs laser treatment, i.e., the global appearance of the non-enveloped ΦX174 virus does not remains intact after treatment which is in agreement with previously established data. [6]

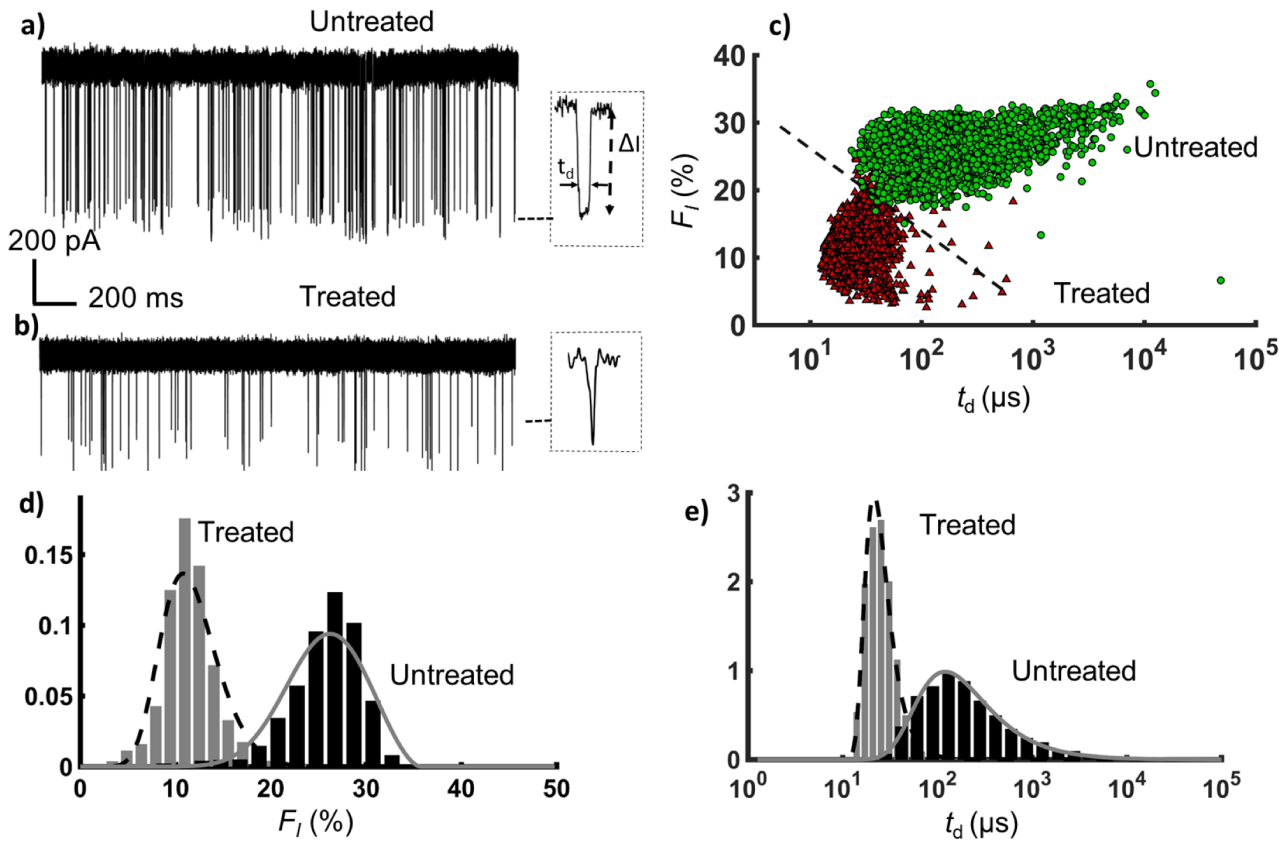

FIG. 3. Comparison of pre-treatment and post-treatment of ΦX174 virus with a pulsed (35 fs) 800&400 nm laser with an average power of 2.5 W for 15 min. a,b) Continuous ion current traces collected using a 40 mv applied voltage low-pass filtered to 100 kHz, Inset: Zoomed=in portion of selected representative events. c) Scatter plot of fractional current blockade percentage, $F_I$(%) at 40 mV (0.1 M KCl, 20°C) vs dwell time, $t_d$. d) Histograms of fractional current blockade at 40 mv showing decrease in $F_I$ (%) after laser treatment. e) Histograms of dwell time at 40 mv voltage demonstrates faster translocation of treated viruses through the pore.

Interestingly using the capability of our label-free resistive pulse technique, one can monitor the inactivation of viruses by looking for presence of intact viruses in the solution with high sensitivity. One simple model is to consider an ellipse in the scatter plot of untreated events centered at the average value of fractional current in y direction and the average of $\log_{10}(t_d)$ in x direction. The semi-minor and semi-major axis radius of this ellipse is equal to three times the standard deviation of the $F_I$ and $\log_{10}(t_d)$ of untreated events, respectively. (Fig. 4a). The standard deviations are obtained based on the Gaussian fits of the corresponding data. Counting the number of the treated (red) data points in the black ellipse gives us a rough estimate of the number of intact viruses in the treated sample, $n$. The number of points in untreated and treated scatter plot is 2,723 and 2,564, respectively. The ratio of the $n$ to the total number of red data points of the treated sample, $N$, gives us an estimation of the survival fraction ($r$)

$$r = \frac{f_I}{f_T} \frac{n}{N} \quad (4)$$

where $f_I$ and $f_T$ are the capture probabilities of intact and treated virus. In our case by assuming these probabilities to be equal, the survival fraction is approximately 0.02. This quantity can provide a simple method for rapidly estimating the efficiency of the inactivation technique.

A more precise way to estimate the $r$ using nanopore data is to determine an upper bound for the number of survived viruses after treatment. To find the upper bound we developed a statistical formula that made use of a probability distribution function derived from Fig. 3d. The first step is to estimate the probability of finding a translocation event of the treated sample with a distinguishing feature $x_T$ inside the untreated sample region with $x_U$. In a simplest case $x_T$ and $x_U$ can be any feature which separates the two populations; in our case it can be fractional current blockade, dwell time or any other combination of these two. For example if we consider these one dimensional scaler as the fractional current blockade, in our case there is a big separation between the mean of untreated events, $\overline{x_U}$, and treated ones, $\overline{x_T}$, so that $\overline{x_T} < \overline{x_U}$. Some of the viruses in the treated sample could be intact. We can quantify this possibility by considering the probability $P_I(x_T)$ that an untreated virus has a value of the feature as low as $x_T$, which is the cumulative distribution function for the untreated sample: $P_I(x_T) = CDF_U(x_T)$. Thus, for the treated data points that are far away from the untreated events, the result would be zero while for the ones which has overlap, the result is more than zero. By assuming all untreated viruses are intact, we can consider a threshold, $\varepsilon$, on the probability $P_I$, and using a Heaviside step function, $\theta$, we decide if an event can be counted as an intact virus (if $P_I(x_T) \geq \varepsilon$) or not (if $P_I(x_T) < \varepsilon$). So for applying this threshold on all events, we can define a function $\rho_\varepsilon$ as equation 5 which is a sum on all treated data points normalized by the number of treated events, $N$.

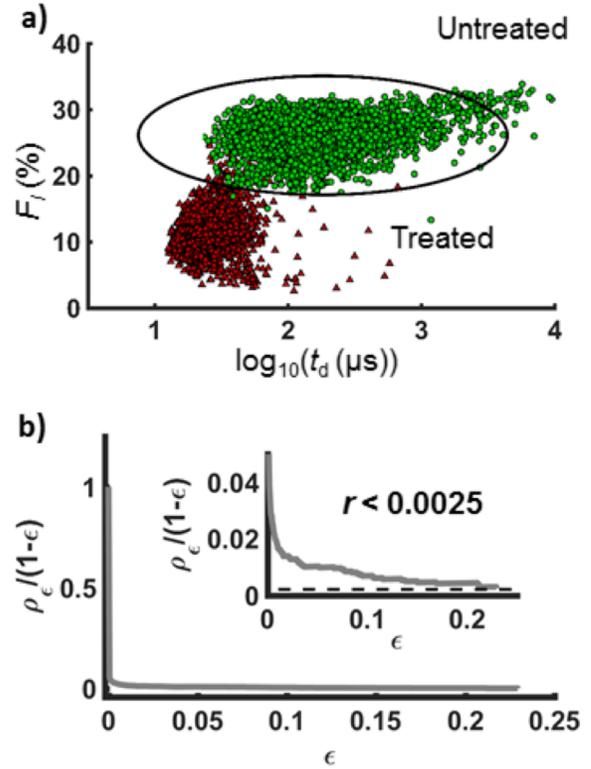

FIG. 4. a) Scatter plot of fractional blockade of untreated (green) and treated (red) viruses versus event duration, under a driving voltage of 40 mV b) upper bound determination of the survival fraction, inset shows the zoom in portion of the graph. The minimum occurs at ε= 0.2294.

$$\rho_\varepsilon = \sum_{i=1}^{N} \frac{\theta(P_I(x_i) - \varepsilon)}{N} \quad (5)$$

So after treatment, there is $\rho_\varepsilon \times N$ number of events in the treated sample which have the possibility bigger than $\varepsilon$ to be counted as the intact viruses. This count can be written as

$$\rho_\varepsilon \times N = n \times (1 - \varepsilon) + (N - n) \times q_\varepsilon \quad (6)$$

Where $n$ in the first part is the actual number of intact viruses in the treated sample, and $q_\varepsilon$ is the possibility of counting a broken virus as an intact one in the treated sample. In this calculation, since $0 \leq q_\varepsilon \leq 1$, the $r$ which is $n/N$, will be bounded by $r \leq \frac{\rho_\varepsilon}{1-\varepsilon}$. By minimizing the $\frac{\rho_\varepsilon}{1-\varepsilon}$ over the $\varepsilon$ one can determine the upper bound for survival fraction, but still there is another point that we need to explore. This value of $r$ has a large error bar due to the shot noise, so that the error in estimating $n$ is proportional to the $\sqrt{n}$. In practice for large enough epsilon there will be no data points satisfying the condition of $P_I(x_T) \geq \varepsilon$ and the estimated value of $\rho_\varepsilon$ would be zero. Thus as a practical approach, we consider a cutoff and vary epsilon from 0 to the maximal value at which $\rho_\varepsilon \times N \geq 5$, where 5 is an arbitrary threshold that controls how accurate this method is. As shown in Fig. 4b in our case, $r_{max}$ can be estimated as $2.5 \times 10^{-3}$ which is similar to the inverse of the reduction value $\sim 10^3$ reported using infectivity assays (Fig. 1b). So these data show that the proposed formulation makes it possible to determine the inactivation efficiency using the nanopore method; while in contrast to the conventional infectivity assays, nanopore technique is capable of probing small damages to the viruses in linear base with high precision.

Generally the electrophoretic driving force and electroosmotic flow are the main effects that govern a particle's motion in a nanopore [43]. The first one originates from the force exerted to the charged nanoparticle in an electric field and the second one is due to the viscous drag by the fluid flowing through the charged nanopore in response to the applied voltage. Combining the equations for electrophoretic and electroosmotic transport yields an effective virus velocity $v$ in an external electric field E inside the pore as [44]

$$v = \frac{E\varepsilon}{\eta}\left(\zeta_{virus} - \zeta_{pore}\right) \quad (7)$$

Where $\zeta$ is zeta potential, $\eta$ and $\varepsilon$ are the viscosity and permittivity of the electrolyte. For our negatively charged viruses attempting to translocate through the nanopore with negative surface charge, these two forces oppose each other and the relative magnitudes of the forces determines the direction and duration of translocations [45].

In our study, electroosmotic flow is a dominant factor, and therefore, the difference in membrane and viruses surface charge plays a dominant role on capture and transport. As shown in Fig. 3e, the dwell time of most of the events in the treated sample are <90 $\mu$s, while untreated viruses have much longer dwell times. In addition, upon treatment, the dwell time distribution narrows substantially and the fractional current blockades are reduced, which can be attributed to a morphological change in the virus sample.

To further examine the impact of fs laser on virus conformation, we employed DLS to characterize changes in the radius of ΦX174 viruses and its DNA caused by laser treatment. The DLS measured the average diameter of the untreated viruses 30.9±2.3 nm (Fig. 5a), which is in agreement with the ΦX174 size.[46] DLS data entails two useful pieces of information: first it shows that the laser-treated sample no longer possesses a monomodal size distribution and a second peak is observed at 458±56 nm which will be discussed shortly. Another important piece of information can be obtained from the intersection of the correlation curve on the y-axis of the correlogram. This y-intercept is related to the signal-to-noise ratio (SNR) of the measured sample. As shown in Fig. 5b the correlation coefficient for the treated ΦX174 viruses has lower y-intercept of 0.4 compared to the untreated one which is ~0.85. The lower y-intercept in the treated sample can be attributed to the concentration variations in this sample and presence of 458 nm particles. This concentration variation can also be explored by nanopore measurements. In case of the nanopore measurements, the waiting time between two successive events (*δt*) is inversely proportional to the concentration of the detectable particles. An exponential fit to the waiting time data (Fig. 5c), indicates that capture rate of the viruses has decreased by more than one order of magnitude after laser treatment at the same voltage bias (40 mV). Capture rates for untreated and treated samples were 52±0.87 s$^{-1}$ and 3.49±0.13 s$^{-1}$ respectively.

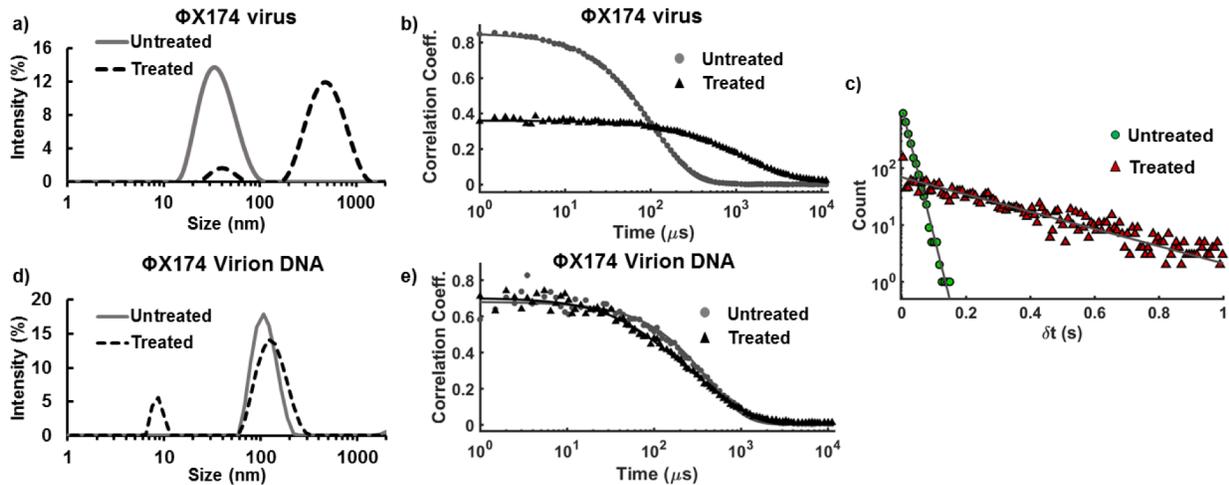

FIG. 5. a) DLS size distribution obtained from b) Quasi-elastic light scattering intensity autocorrelation function or correlogram of ΦX174 virus before (untreated) and after irradiation (treated) with IR& blue 35 fs laser pulses for 30 min (average laser power 2.5 W) c) inter-event waiting time obtained from nanopore measurement of the data in Fig. 3 at 40 mv applied bias, fitted by an exponential function. d) Size distribution e) Correlogram of ΦX174 viral DNA. Laser irradiation conditions were identical to (a,b).

Particles of the second mode in the DLS data of ΦX174 viruses have a hydrodynamic radius that is too large to be detected with our 38 nm nanopore. To investigate the second peak, we performed DLS measurement of ΦX174 virion DNA, the single-stranded viral DNA isolated from purified phage by phenol extraction (New England Biolabs). This measurement indicates weather the second peak is a signature of free DNA ejected from virus under photonic exposure. The DLS measured average diameter of the untreated phenol-extracted DNA was 113.5±30 nm. A part of this variation may be attributed to ~ 15% of the phenol-extracted DNA molecules not being in circular form. The radius of gyration, $R_g$ and hydrodynamic radius, $R_h$ of free ΦX174 DNA polymer can be estimated from the known persistence length

$$R_g = \sqrt{\frac{2PL_c}{6}} \qquad (8)$$

where *P* is persistence length and $L_c$ is contour length. [47-49] For ΦX174 DNA with 5386 nucleotides and the persistence length ~4.6 nm for DNA, [50] the $R_g$ and $R_h$ can be estimated as 72 nm and 45.6 nm, respectively. [41,51] So the corresponding hydrodynamic diameter is approximately 91 nm which is consistent with the peak in DLS measurement of untreated DNA. Interestingly, DLS data of laser exposed DNA (Fig. 5d), still contains this peak which argues against complete agglomeration of single-stranded DNA due to laser exposure. Also, as the correlogram of the DNA (Fig. 5e) shows there is little difference in the time autocorrelation function between exposed and unexposed DNA samples. Our observations indicate that the 458 nm peak observed in laser-treated ΦX174 viruses sample is not from the presence of free single-stranded DNA.

Based on our data after laser treatment there are three possible morphology changes that can happen to ΦX174 viruses: (i) fragmentation to small pieces, (ii) perforation of viruses that leads to DNA expulsion with a small associated change in the virus shell, and (iii) agglomeration of viral particles (Fig. 6). Nanopore data can be used to evaluate each of the possibilities. The fact that the fractional current blockade decreases after treatment, suggests that the particle diameter gets smaller upon laser exposure. Also if we neglect possible changes in the zeta potential of the viruses, the smaller viruses will translocate faster consistent with the observation of a shorter dwell times in the treated sample. Therefore, the current blockade and the dwell time both point to detection of entities smaller than the original viruses, and support the hypothesis of virus perforation. At the same time, the nanopore capture rate suggests that viruses for the most part have been fragmented to small pieces not detectable by the nanopore conductance.

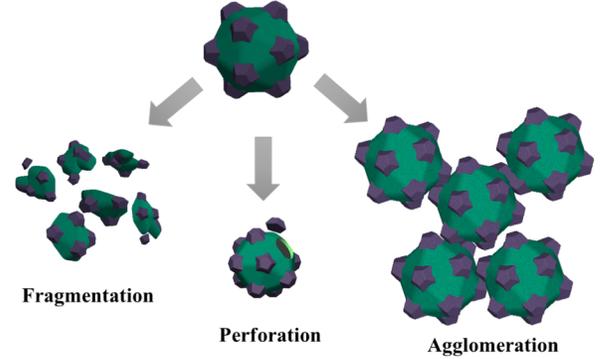

FIG. 6. Schematic of the possible effect of the fs laser treatment on ΦX174 virus.

Neither DLS nor nanopore data don't suggest extensive virus agglomeration. Since the differential scattering cross-section for elastic scattering in the Rayleigh regime is proportional to the 6$^{th}$ power of the particle radius with given dielectric properties, DLS may be expected to be sensitive to agglomeration or presence of larger particle species. Additionally, significant agglomeration is expected to lead to a higher intensity at second diffraction mode which is not observed. Also in nanopore data, significant agglomeration of viruses can block the pore which leads to long lasting dwell times in contrast to what is observed in our experiments. So taken together, our data suggest that a small fraction of viruses (7%) get perforated, while a negligible fraction (< 10$^{-3}$% based on Rayleigh scattering assumption) get agglomerated, and the remaining largest fraction of the viruses (93%) are fragmented.

While DLS provides an ensemble averaged estimate of the particle size distribution, it does not provide information at the single particle level. Our experiments suggest nanopore single-particle conductance measurements to be helpful in detecting intact viruses in small-volume samples (∼0.5 μl) to characterize morphological changes caused by laser treatment and inactivation.

## IV. CONCLUSION

The ultrashort pulsed laser technology presented here can be readily used for the rapid and effective disinfection of viruses in a label free manner. The application of this technology to the disinfection of viruses can lead to some changes in the viruses which needs to be monitored with high precision. Nanopore technique allowed us to detect intact viruses and monitor the effect of fs laser on a model virus in a single molecule level. Analysis of changes in the ionic current through the nanopore provides information about size and physical properties of viruses before and after laser treatment. To the best of our knowledge, this is the first time that nanopores have been used to monitor changes induced by an inactivation method in viruses.

Our data is a promising step towards developing a label-free detection technique that can also be used as an effective method for monitoring the survival fraction of viruses using low sample volume, high precision and fast assay time.

## ACKNOWLEDGMENTS

Grants: Kirill S. Korolev was partially supported by Cottrell Scholar Award (#24010) by the Research Corporation for Science Advancement, a grant from the Simons Foundation (#409704) and an award from Gordon and Betty Moore foundation (#6790.08). X.L. and H.Y. were supported by the China Scholarship Council (CSC), and M.A.A. was supported by the National Institutes of Health (R01-HG009186).

[1] P. Hellstern, Current opinion in hematology **11**, 346 (2004).
[2] G. Rock, Vox sanguinis **100**, 169 (2011).
[3] J. AuBuchon, ISBT Science series **6**, 181 (2011).
[4] K.-T. Tsen, S.-W. D. Tsen, C.-L. Chang, C.-F. Hung, T.-C. Wu, and J. G. Kiang, Journal of biomedical optics **12**, 064030 (2007).
[5] S.-W. D. Tsen, T. Chapa, W. Beatty, K.-T. Tsen, D. Yu, and S. Achilefu, Journal of biomedical optics **17**, 128002 (2012).
[6] S.-W. D. Tsen, D. H. Kingsley, C. Poweleit, S. Achilefu, D. S. Soroka, T. Wu, and K.-T. Tsen, Virology journal **11**, 20 (2014).
[7] K. T. Tsen, S.-W. D. Tsen, Q. Fu, S. M. Lindsay, Z. Li, S. Cope, S. Vaiana, and J. G. Kiang, Journal of biomedical optics **16**, 078003 (2011).
[8] M. Nazari *et al.*, Scientific reports **7**, 11951 (2017).
[9] S.-W. D. Tsen, Y.-S. D. Tsen, K. Tsen, and T. Wu, Journal of Healthcare Engineering **1**, 185 (2010).
[10] K. Tsen, S.-W. D. Tsen, C.-F. Hung, T. Wu, and J. G. Kiang, Journal of Physics: Condensed Matter **20**, 252205 (2008).
[11] H. O. Kangro and B. W. Mahy, *Virology methods manual* (Academic Press, 1996).
[12] G. Daaboul, A. Yurt, X. Zhang, G. Hwang, B. Goldberg, and M. Unlu, Nano letters **10**, 4727 (2010).
[13] A. Mitra, B. Deutsch, F. Ignatovich, C. Dykes, and L. Novotny, ACS nano **4**, 1305 (2010).
[14] C. L. Stoffel, R. F. Kathy, and K. L. Rowlen, Cytometry Part A **65**, 140 (2005).
[15] A. Piruska, M. Gong, J. V. Sweedler, and P. W. Bohn, Chemical Society Reviews **39**, 1060 (2010).
[16] Z. D. Harms, D. G. Haywood, A. R. Kneller, L. Selzer, A. Zlotnick, and S. C. Jacobson, Analytical chemistry **87**, 699 (2014).
[17] B. Ledden, D. Fologea, D. S. Talaga, and J. Li, in *Nanopores* (Springer, 2011), pp. 129.
[18] J. Li, D. Fologea, R. Rollings, and B. Ledden, Protein and peptide letters **21**, 256 (2014).
[19] A. Oukhaled *et al.*, ACS nano **5**, 3628 (2011).
[20] D. S. Talaga and J. Li, Journal of the American Chemical Society **131**, 9287 (2009).
[21] K. J. Freedman, S. R. Haq, J. B. Edel, P. Jemth, and M. J. Kim, Scientific reports **3** (2013).
[22] K. J. Freedman, M. Jürgens, A. Prabhu, C. W. Ahn, P. Jemth, J. B. Edel, and M. J. Kim, Analytical chemistry **83**, 5137 (2011).
[23] B. Cressiot, A. Oukhaled, G. Patriarche, M. Pastoriza-Gallego, J.-M. Betton, L. c. Auvray, M. Muthukumar, L. Bacri, and J. Pelta, ACS nano **6**, 6236 (2012).
[24] D. Branton *et al.*, Nature biotechnology **26**, 1146 (2008).
[25] S. Carson and M. Wanunu, Nanotechnology **26**, 074004 (2015).
[26] M. A. Alibakhshi, J. R. Halman, J. Wilson, A. Aksimentiev, K. A. Afonin, and M. Wanunu, ACS nano **11**, 9701 (2017).
[27] L. Movileanu, S. Cheley, and H. Bayley, Biophysical journal **85**, 897 (2003).
[28] O. Braha, L.-Q. Gu, L. Zhou, X. Lu, S. Cheley, and H. Bayley, Nature biotechnology **18**, 1005 (2000).
[29] K. Zhou, L. Li, Z. Tan, A. Zlotnick, and S. C. Jacobson, Journal of the American Chemical Society **133**, 1618 (2011).
[30] Z. D. Harms *et al.*, Analytical chemistry **83**, 9573 (2011).
[31] N. Arjmandi, W. Van Roy, and L. Lagae, Analytical chemistry **86**, 4637 (2014).
[32] N. Arjmandi, W. Van Roy, L. Lagae, and G. Borghs, Analytical chemistry **84**, 8490 (2012).
[33] H. Wu, Y. Chen, Q. Zhou, R. Wang, B. Xia, D. Ma, K. Luo, and Q. Liu, Anal. Chem **88**, 2502 (2016).
[34] K.-T. Tsen *et al.*, Journal of biomedical optics **14**, 064042 (2009).
[35] S. S. Thompson and M. V. Yates, Applied and environmental microbiology **65**, 1186 (1999).
[36] X. Liu, Q. Dai, L. Austin, J. Coutts, G. Knowles, J. Zou, H. Chen, and Q. Huo, Journal of the American Chemical Society **130**, 2780 (2008).
[37] K.-T. Tsen, S.-W. D. Tsen, C.-L. Chang, C.-F. Hung, T. Wu, and J. G. Kiang, Virology journal **4**, 50 (2007).
[38] S.-W. D. Tsen, T. C. Wu, J. G. Kiang, and K.-T. Tsen, Journal of biomedical science **19**, 62 (2012).
[39] H. Brough, C. Antoniou, J. Carter, J. Jakubik, Y. Xu, and H. Lutz, Biotechnology progress **18**, 782 (2002).

[40] M. J. Kim, M. Wanunu, D. C. Bell, and A. Meller, Advanced materials **18**, 3149 (2006).
[41] P. Waduge, R. Hu, P. Bandarkar, H. Yamazaki, B. Cressiot, Q. Zhao, P. C. Whitford, and M. Wanunu, ACS nano **11**, 5706 (2017).
[42] J. Larkin, R. Y. Henley, M. Muthukumar, J. K. Rosenstein, and M. Wanunu, Biophysical journal **106**, 696 (2014).
[43] G. Huang, K. Willems, M. Soskine, C. Wloka, and G. Maglia, Nature Communications **8**, 935 (2017).
[44] M. Firnkes, D. Pedone, J. Knezevic, M. Döblinger, and U. Rant, Nano letters **10**, 2162 (2010).
[45] D. V. Melnikov, Z. K. Hulings, and M. E. Gracheva, Physical Review E **95**, 063105 (2017).
[46] M. Bayer and R. DeBlois, Journal of virology **14**, 975 (1974).
[47] D. Boal and D. H. Boal, *Mechanics of the Cell* (Cambridge University Press, 2012).
[48] A. Y. Sim, J. Lipfert, D. Herschlag, and S. Doniach, Physical Review E **86**, 021901 (2012).
[49] K. Rechendorff, G. Witz, J. Adamcik, and G. Dietler, The Journal of chemical physics **131**, 09B604 (2009).
[50] G. S. Manning, Biophysical journal **91**, 3607 (2006).
[51] D. R. Tree, A. Muralidhar, P. S. Doyle, and K. D. Dorfman, Macromolecules **46**, 8369 (2013).